\documentclass[twocolumn]{aastex62}
\pdfoutput=1 
\usepackage{amsmath,amstext}
\usepackage[T1]{fontenc}
\usepackage{apjfonts}
\usepackage[figure,figure*]{hypcap}
\usepackage{rotating}
\usepackage{lineno}
\usepackage{hyperref}
\graphicspath{{./}{figures/}}

\definecolor{chmagenta}{rgb}{0.54, 0.17, 0.88}

\shorttitle{Binary BBH Interactions}
\shortauthors{Zevin et al.}

\begin{document}

\title{Eccentric Black Hole Mergers in Dense Star Clusters: \\The Role of Binary-Binary Encounters}

\author{Michael Zevin}\thanks{zevin@u.northwestern.edu}
\affiliation{Department of Physics and Astronomy, Northwestern University, 2145 Sheridan Road, Evanston, IL 60208, USA}
\affiliation{Center for Interdisciplinary Exploration and Research in Astrophysics (CIERA), 2145 Sheridan Road, Evanston, IL 60208, USA}

\author{Johan Samsing}
\affiliation{Department of Astrophysical Sciences, Princeton University, Peyton Hall, 4 Ivy Lane, Princeton, NJ 08544, USA}

\author{Carl Rodriguez}\thanks{Pappalardo Fellow}
\affiliation{Pappalardo Fellow; MIT-Kavli Institute for Astrophysics and Space Research, 77 Massachusetts Avenue, 37-664H, Cambridge, MA 02139, USA}

\author{Carl-Johan Haster}
\affiliation{Canadian Institute for Theoretical Astrophysics, 60 St. George Street, Toronto, Ontario, M5S 3H8, Canada}
\affiliation{LIGO Laboratory and MIT-Kavli Institute for Astrophysics and Space Research, 77 Massachusetts Avenue, 37-664H, Cambridge, MA 02139, USA}

\author{Enrico Ramirez-Ruiz}
\affiliation{Department of Astronomy and Astrophysics, University of California, Santa Cruz, California 95064, USA}
\affiliation{Niels Bohr Institute, Blegdamsvej 17, 2100 K\o benhavn \O, Denmark}

\begin{abstract}
We present the first systematic study of strong binary-single and binary-binary black hole (BH) interactions with the inclusion of general relativity. 
By including general relativistic effects in the equations of motion during strong encounters, the dissipation of orbital energy from the emission of gravitational waves (GWs) can lead to captures and subsequent inspirals with appreciable eccentricities when entering the sensitive frequency ranges of the LIGO and Virgo GW detectors. 
It has been shown that binary-single interactions significantly contribute to the rate of eccentric mergers, but no studies have looked exclusively into the contribution from binary-binary interactions. 
To this end, we perform binary-binary and binary-single scattering experiments with general relativistic dynamics up through the 2.5 post-Newtonian order included, both in a controlled setting to gauge the importance of non-dissipative post-Newtonian terms and derive scaling relations for the cross section of GW captures, as well as experiments tuned to the strong interactions from state-of-the art globular cluster (GC) models to assess the relative importance of the binary-binary channel in facilitating GW captures and the resultant eccentricity distributions of inspiral from channel. 
Although binary-binary interactions are 10--100 times less frequent in GCs than binary-single interactions, their longer lifetime and more complex dynamics leads to a higher probability for GW captures to occur during the encounter. 
We find that binary-binary interactions contribute 25--45\% of the eccentric mergers that occur during strong BH encounters in GCs, regardless of the properties of the cluster environment. 
The inclusion of higher multiplicity encounters in dense star clusters therefore have major implications on the predicted rates of highly eccentric binaries potentially detectable by the LIGO/Virgo network. 
Becse gravitational waveforms of eccentric inspirals are distinct from those generated by merging binaries that have circularized, measurements of eccentricity in such systems would highly constrain their formation scenario.

\end{abstract}


\section{Introduction}

The multiple discoveries of coalescing binary black hole (BBH) systems by the advanced network of gravitational-wave (GW) interferometers \citep{GW150914,GW151226,O1_BBH,GW170104,GW170608,GW170814} has led to significant interest in the astrophysical mechanisms responsible for their formation and subsequent merger. 
One evolutionary channel that may contribute greatly to the population of BBHs is dynamical formation within dense stellar environments such as globular clusters (GCs) and nuclear star clusters (NSCs) \citep{PortegiesZwart2000,Downing2009,Downing2011,Rodriguez2015,Rodriguez2016a,Fragione2018}, as well as young massive and open clusters \citep{Banerjee2017,Banerjee2018a,Banerjee2018b}.
Through dynamical friction, black holes (BHs) tend to migrate toward the cores of clusters, where stellar densities can be over a million times higher than that of our solar neighborhood \citep{lightman1978}. 
In these tightly packed collisional environments, BHs frequently interact with one another, swapping partners and hardening their orbits, and thereby losing any memory of their primordial orbital states \citep[e.g.,][]{McMillan1991,Hut1992,Fregeau2007}.  
BBH mergers from dynamical environments thus imprint unique and potentially detectable characteristics in their GW waveforms relative to BBHs whose progenitors evolved in isolation, providing a possible route for discriminating between the various scenarios proposed for BBH formation. 

In recent years, much attention has been focused on the BH spin orientations as a means to discriminate different BBH formation channels \citep{Rodriguez2016,Stevenson2017,Talbot2017,Vitale2017a,Farr2017,Farr2017a,Gerosa2018a,Sedda2018,Schrøder2018}; BBHs that evolve in isolation are expected to have spin vectors that are relatively aligned with the angular momentum of the binary, whereas BBHs that assemble dynamically will have spin vectors distributed isotropically on the sphere. 
However, if the spin magnitudes of heavy BHs are naturally low, the ability to discern formation scenarios using spin parameters is stifled. 
Mass distributions may also be useful once dozens to hundreds of observations are made \citep{Stevenson2015,Mandel2016a,Zevin2017b} or if second-generation BH mergers are found with masses in the putative pair instability upper mass gap \citep{OLeary2016,Fishbach2017,Rodriguez2018b}.

While our ability to measure BBH spins may be stymied by low spin magnitudes, the orbital eccentricity of the binary is entirely a function of the well-understood dynamics that assembled the system. 
Eccentricity is often overlooked when discussing the parameters of merging BBHs; GW emission is highly efficient at circularizing the orbit of an inspiraling binary \citep{Peters1964} and most formation scenarios predict the binary to have evolved in isolation for substantial periods of time before the merger, thereby circularizing its orbit to a point where any measurable semblance of eccentricity would be lost before entering the sensitive frequency band of ground-based GW detectors. 
Furthermore, matched-filtering searches for GWs do not utilize eccentric templates \citep{Usman2016,Messick2017}, necessitating methods of detection that are promising but significantly less effective than matched-filtering \citep{Tai2014,Coughlin2015,Tiwari2016,Huerta2017,Gondan2017,Gondan2017a,Huerta2018,Klein2018,Rebei2018,Gondan2018}.
However, recent work modeling the strong binary-single encounters that harden BBHs in GCs find that resonating interactions (RIs) of BH systems (i.e., interactions of three or more bodies that evolve chaotically over many orbital times before dissociating) can facilitate rapid and highly eccentric mergers when post-Newtonian (pN) effects, particularly the emission of GWs, are included \citep{Gultekin2006,Samsing2014,Haster2016,Samsing2017c,Rodriguez2018b,Samsing2018d,Samsing2018c, Banerjee2018b}.
Though the modeling of chaotic BH interactions has traditionally relied on Newtonian $N$-body simulations \citep[e.g.,][]{Hut1983,Fregeau2004}, it was shown semi-analytically in \cite{Samsing2014, Samsing2017e} and  \cite{Samsing2018d}, and with full numerical simulations in \cite{Rodriguez2018b}, that the inclusion of pN terms in the $N$-body equations of motion has a significant impact on the evolution and outcome of such encounters. 

During RIs, numerous meta-stable \textit{intermediate-state} (IMS) binaries form before the interaction ceases through the ejection of enough components. 
These encounters can be long-lived, especially when the mass ratios are near unity \citep{Sigurdsson1993}, thus leading to dozens of IMSs during a single RI. 
Each IMS binary synthesized during the interaction will acquire an orbital eccentricity drawn from a quasi-thermal distribution. 
If an IMS binary has a high enough eccentricity (or if two unbound compact objects pass close enough to one another during such an encounter), gravitational radiation will significantly dissipate orbital energy during periapse passages, which can lead to a GW capture and rapid inspiral \citep{Quinlan1987}. 
Due to the swiftness of these inspiral timescales, the system will not have time to fully damp its orbital eccentricity, leading to appreciable eccentricities even in the frequency ranges of ground-based GW interferometers such as the Advanced LIGO \citep{aLIGO} and Advanced Virgo \citep{aVirgo} detectors, with as many as $\sim$\,5\% of these events having eccentricities greater than $0.1$ at a GW frequency of 10 Hz \cite[e.g.,][]{Samsing2018d,Rodriguez2018b}. 

Other formation scenarios that may facilitate BBH inspirals with eccentricities accessible by current ground-based GW detectors have also been identified, including dynamical interactions with an intermediate-mass BH in a GC core \citep{Leigh2014,Fragione2017,Fragione2018a}, hyperbolic encounters between BHs in NSCs \citep{OLeary2009,Kocsis2012}, and through the evolution of three- and four-body hierarchical systems \citep[e.g., ][]{Miller2002,Antonini2012,Antonini2016a,Silsbee2017,Antonini2017a,Rodriguez2018,Hoang2017,Liu2017,Liu2018,Liu2018a,Randall2018,Randall2018b,Hamers2018,Arca-Sedda2018}, particularly when non-secular evolution is properly considered \citep{Antonini2014}.
Motivated by formation scenarios with this highly discriminating characteristic, efforts have been made to quantify the measurability of eccentric signals, and have placed limits on the amount of eccentricity needed in a signal to distinguish it from circular. 
For example, \cite{Lower2018} find that eccentricities will be discernible for a signal analogous to GW150914 \citep{GW150914} detected by the Advanced LIGO/Virgo network if the eccentricity is $\gtrsim$\,0.05 at a GW frequency of 10 Hz. 

Recent work modeling BH encounters in GCs with pN dynamics have focused on binary-single BH encounters. 
In this paper, we present the first systematic study of binary-binary BH scattering encounters with pN terms up to and including the 2.5pN order.
In addition to gauging the dependence of the GW capture cross section on initial conditions of the binary-binary configuration, we use binary-single and binary-binary interactions from state-of-the-art cluster models to compare the efficiency and relative rate of captures and inspirals from these two types of encounters. 
We find that, while binary-single encounters are more than an order of magnitude more prevalent in cluster environments, binary-binary encounters are naturally more efficient at inducing GW captures during an RI; in total, binary-binary interactions contribute 25--45\% of highly eccentric GW inspirals in GCs, irrespective of the cluster properties.
Similar to binary-single interactions in GCs, BBH mergers from binary-binary encounters lead to three distinct populations of eccentric GW inspirals \citep{Rodriguez2018b,Samsing2018}.
Though only the most rapidly inspiraling population has eccentricities accessible by ground-based GW detectors, eccentricity measurements of the other two populations will be attainable by future space-based interferometers such as LISA \citep{Samsing2018,DOrazio2018}.

We first outline the numerical methods and pN additions to the $N$-body equations of motion in Section \ref{sec:numerical_methods}. 
In Section \ref{sec:scattering_experiments}, we derive analytical approximations and perform scattering experiments to investigate the post-encounter orbital properties from binary-binary interactions (Section \ref{subsec:orbital_properties}). 
We then discuss the dependence of various outcomes, including GW captures, on the initial properties of binary-binary systems, and for the first time quantify how non-dissipative pN terms affect the probability of GW captures in such encounters (Section \ref{subsec:fiducial_strong_encounters}). 
Following this, we gauge the relative efficiency of inducing GW captures from binary-binary encounters, compared to their binary-single counterparts, using state-of-the-art cluster models, and examine the eccentricity distribution of inspirals from binary-binary encounters (Section \ref{subsec:GC_encounters}). 
We discuss the implications of our findings and future work in Section \ref{sec:discussion}, and summarize our main conclusions in Section \ref{sec:conclusions}.

\section{Numerical methods}\label{sec:numerical_methods}

Orbital dynamics involving more than two bodies is chaotic; no general analytic solution can be derived and subtle changes in the initial conditions of the system can lead to vastly different outcomes \citep[e.g.,][]{Samsing2018e}. 
Therefore, it is common to perform a large number of scattering experiments that span the possible initial configurations in order to quantitatively determine how variations in initial conditions probabilistically affect interaction outcomes. \citep{Heggie1975,Hut1983,Fregeau2004,Antognini2016}.
In particular, three-body binary-single scatterings have been extensively studied \citep[e.g.,][]{Fregeau2004}, and more recently this problem has been reexamined with the inclusion of GW dissipation in the equations of motion \citep{Samsing2014}. 
The problem of four-body binary-binary scattering has been investigated to a lesser extent, as the multitude of possible final configurations and necessary computational requirements make higher multiplicity encounters much more complicated to examine with scattering experiments. 
However, cluster modeling predicts that a significant number of binary-binary BH encounters do occur in the cluster cores \citep{Antonini2016a}, and such encounters are vital for the formation of triple systems because the Newtonian energetics of three-body encounters in GCs will typically not allow for the formation of a bound triple. Binary-binary interactions have also been shown to amplify the Lidov-Kozai mechanism in BBH systems \citep{Miller2002,Liu2018a} as well as provide a potential explanation for abnormal pulsar accelerations in certain GCs \citep{Colpi2003}. 

In investigating the dynamical impact of binary-binary encounters, studies such as \cite{Fregeau2004} and later \cite{Antognini2016} performed detailed scattering experiments for binary-binary interactions \cite[as well as other higher-multiplicity systems in][]{Antognini2016} in the Newtonian regime to comprehensively gauge how variations in initial conditions affect the cross sections of particular outcomes. 
However, as these studies did not specifically target encounters of compact objects or account for pN effects, there have been no studies that investigate the role of binary-binary encounters involving BH systems in the strong gravity regime and how such interactions instigate GW captures and rapid inspirals. 

In this study, we perform $O(10^5)$ scatterings for each set of initial conditions to determine cross sections of particular outcomes, as well as properties of the BBH inspirals that follow.

\begin{figure*}[t!]
\includegraphics[width=1.0\textwidth]{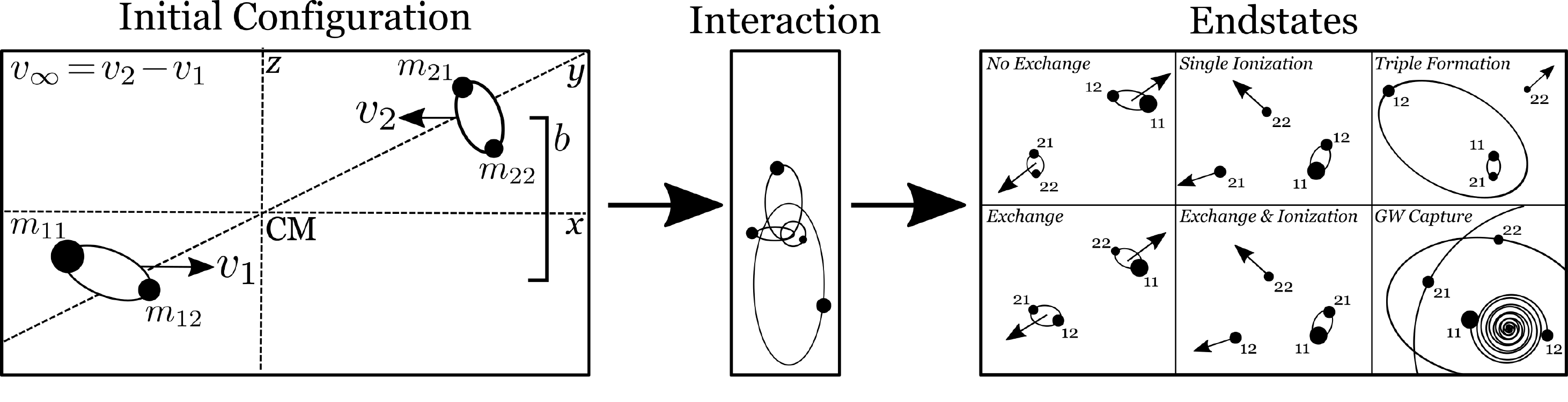}
\caption{Schematic of initial configuration, resonant interaction, and potential endstates of a pN binary-binary BH encounter. }
\label{fig:schematic}
\end{figure*}

\subsection{Initial Conditions and  Pre-Encounter setup}\label{subsec:initial_conditions}

Each binary-binary system is defined by its four component masses, two semi-major axes (SMAs), and two eccentricities prior to interaction, as well as the relative incoming velocity of the two pre-encounter binaries. 
Masses and orbital parameters are notated by subscripts in a top-down fashion as in \cite{Fregeau2004}, where leftmost indices in the subscripts denote the separate binary systems prior to interaction and rightmost indices the components of the binary. 
For example, $m_{12}$ indicates the secondary component of the target binary and $a_{2}$ the SMA of the incoming binary. 
We sample the location of the orbit by solving Kepler's equations numerically and sampling the mean anomaly. We then randomly sample the three orientation angles of each binary: $\phi_{\rm peri} = [0,2\pi],\ \cos(\theta_{\rm i}) = [-1,1],\ \phi_{\rm ascn} = [0,2\pi]$ where $\phi_{\rm peri}$ is the angle of periapse, $\theta_{\rm i}$ is the inclination, and $\phi_{\rm ascn}$ is the angle of ascending node. 

Given an incoming velocity at infinity $v_{\infty}$, we define our maximum impact parameter similarly to \cite{Hut1983}\footnote{The maximum sampled impact parameter is only used for gridded scattering experiments; for encounters that we extract from cluster models, the true impact parameter is recorded and used.}: 

\begin{equation}
b_{\rm max} = \left(\frac{4 v_{\rm crit}}{v_{\infty}} + 3\right) a_{\rm max}
\label{eq:inpact_parameter}
\end{equation}
where $v_{\rm crit}$ is the critical velocity at which the total energy of the system is zero and $a_{\rm max}$ is the largest of the two binary SMAs.
We then draw the impact parameter of the incoming system at infinity ($b_{\infty}$) uniformly from a circle of radius $b_{\rm max}$. 
To limit integration time at large separations, we analytically evolve the incoming system forward from $b_{\infty}$ and $v_{\infty}$ using conservation of energy and angular momentum until one of the two binary systems reaches a threshold point of $F_{\rm tid}/F_{\rm rel} = 10^{-5}$, where $F_{\rm tid}$ is the tidal force on the components of one binary from the other binary and $F_{\rm rel}$ is the gravitational force between two components within a single binary. 
We then integrate the pN equations of motion until a physical or computational outcome is reached using the $N$-body integration scheme detailed in \cite{Samsing2017d}.
As we only consider the scattering of systems composed entirely of BHs, finite-size effects such as tides are ignored.

\subsection{Quantifying Interaction Probability}

Binary-binary scattering experiments lead to various potential outcomes, which we refer to as \textit{endstates}. 
To quantify the probability of a particular endstate, we define a cross section in the standard way: 
\begin{equation}
\sigma_{\rm X} = \pi b^{2} \frac{N_{\rm X}}{N_{\rm tot}}
\end{equation}
where $N_{\rm X}$ is the number of realizations that result in endstate $X$ for a given initial condition and $N_{\rm tot}$ is the total number of realizations run for a particular initial condition. 
From this, the rate for a particular outcome is approximated as $\Gamma_{\rm X} \simeq n_{\rm BBH} \sigma_{\rm X} v_{\rm disp}$ where $v_{\rm disp}$ is the velocity dispersion and $n_{\rm BBH}$ is the number density of BBHs in the cluster core.
When investigating the relative rate of a particular outcome, it is useful to normalize the cross section by the sum of the areas of the two interacting binaries, a quantity referred to as the reduced cross section: 
\begin{equation}
\hat{\sigma}_{\rm X} = \frac{b^{2}}{a_{1}^2 + a_{2}^2} \frac{N_{\rm X}}{N_{\rm tot}}.
\end{equation}

There are two types of uncertainty to consider in our scattering experiments. 
The first is statistical uncertainty due to the finite number of scattering experiments, which is simply a Poisson counting uncertainty: 
\begin{equation}
\Sigma_{\rm X,stat} = \frac{\sigma_{\rm X}}{\sqrt{N_{\rm X}}}. 
\end{equation}
The second source of uncertainty is due to computational constraints; certain interactions will form long-lived metastable states or be thrown into wide orbits, which take an exceedingly long time to integrate. 
We mark systems as \textit{unresolved} if they simulate for longer than $10^{4}$ times the average initial orbital time of the two incoming binaries or if the computing time of the integration exceeds 1\,hr. 
Given $N_{\rm unres}$ unresolved systems, the resultant systematic uncertainty is 
\begin{equation}
\Sigma_{\rm X,sys} = \pi b_{\rm max}^2 \frac{N_{\rm unres}}{N_{\rm tot}},
\end{equation}
which only acts to increase the cross-section uncertainty. 

We typically find that $\lesssim$\,5\% of systems for a particular initial configuration are unresolved due to integrating for a simulation time of longer than 10$^4$ average initial orbital times, and $\lesssim$\,1\% are unresolved due to exceeding an hour of computer integration time. 
However, these unresolved outcomes still dominate over low-probability endstates such as GW captures.
Throughout this text, our upper error bars show only statistical uncertainty, for readability. 
We find this to be reasonable, as it is expected that, if fully integrated, the endstates for unresolved systems will proportionally follow the endstate cross sections of resolved systems. 
However, we still include the cross sections of unresolved systems as separate points in our figures that, if added to the cross section of another endstate, will provide a highly conservative upper limit. 
We stress again that this uncertainty can only act to \textit{increase} the cross section of resolved endstates.

\subsection{Halting Criteria and Possible Outcomes}\label{subsec:endstates}

We define endstates similar to those in \cite{Antognini2016}, with the addition of the crucial ``GW capture'' endstate, which becomes relevant when pN effects are considered (see Figure \ref{fig:schematic} for schematics of possible endstates). 
In all endstates, we define an object as ``unbound'' when it has positive energy relative to all other components, is moving away from center of mass of the interaction, and its tidal force on other components in the interaction is less than $10^{-3}$ times their relative binding force. The possible endstates are defined as follows:

\begin{itemize}

\item \textsc{No Exchange}: Two bound binaries unbound from one another, with constituent components that are identical to the initial configuration. 
This can result from either a weak interaction fly-by or an RI that leads to a final configuration identical to the initial configuration. 

\item \textsc{Exchange}: Two bound binaries unbound from one another, with constituent components that are different from the initial configuration. 

\item \textsc{Single Ionization}:  One bound binary that maintained its initial configuration and ionized the two components of the other binary. 

\item \textsc{Exchange \& Ionization}: One bound binary composed of two components that originated in different binaries, with the two remaining components ionized. 

\item \textsc{Triple Formation}: One of the four components is ionized and a stable hierarchical triple is formed. 
We determine whether a triple is stable according to the stability criterion from \cite{Mardling2001}: 
\begin{equation}
\frac{a_{1}(1-e_{1})}{a_{11}} > 2.8 \left[ \left(1 +\frac{m_{11}}{m_{12}} \right) \frac{1+e_{1}}{\sqrt{1-e_{1}}} \right]^{2/5} \left( 1-\frac{0.3 i}{\pi} \right)
\end{equation}
where $i$ is the inclination of the outer component's orbit relative to the orbital plane of the inner binary, and once we again use the top-down notation (for example, $a_{11}$ is the SMA of the inner binary in the triple, $m_{12}$ is the total mass of the inner binary, and $m_{11}$ is the mass of the tertiary; see \cite{Fregeau2004}). 
Systems that reach this stable endstate can by examined further through secular evolution.

\item \textsc{GW Capture}: 
Emission of GWs lead to a rapid inspiral and merger of two component BHs during the RI \citep{Samsing2014}. 
To avoid the breakdown of our numerical integration schemes as the two quasi-point particles come near contact, we determine this endstate when two particles are in a bound orbit ($E_{\rm ij} < 0$) and their SMA reaches a nominal value, namely $a_{\rm ij}/(R_{{\rm s},i}+R_{{\rm s},j}) < 10$, where $R_{\rm s}$ is the Schwarzschild radius. 
Systems meeting this criterion will merge on a rapid timescale and perturbations from other component BHs in the RI can be neglected; for example, two 20\,$M_{\odot}$ BHs on a circular orbit at this limit will merge in less than three seconds. 

\end{itemize}

Two additional endstates are possible: a full ionization (i.e., all components of the two binary systems become ionized) and a direct collision. 
However, these endstates are exceedingly rare, relative to the other endstates; because $v_{\infty} \ll v_{\rm crit}$ for most cluster binaries, fully ionizing encounters are energetically improbable, and the physical sizes of stellar-mass BHs make direct collisions in unbound systems highly unlikely.

\begin{figure*}[t!]
\includegraphics[width=1.0\textwidth]{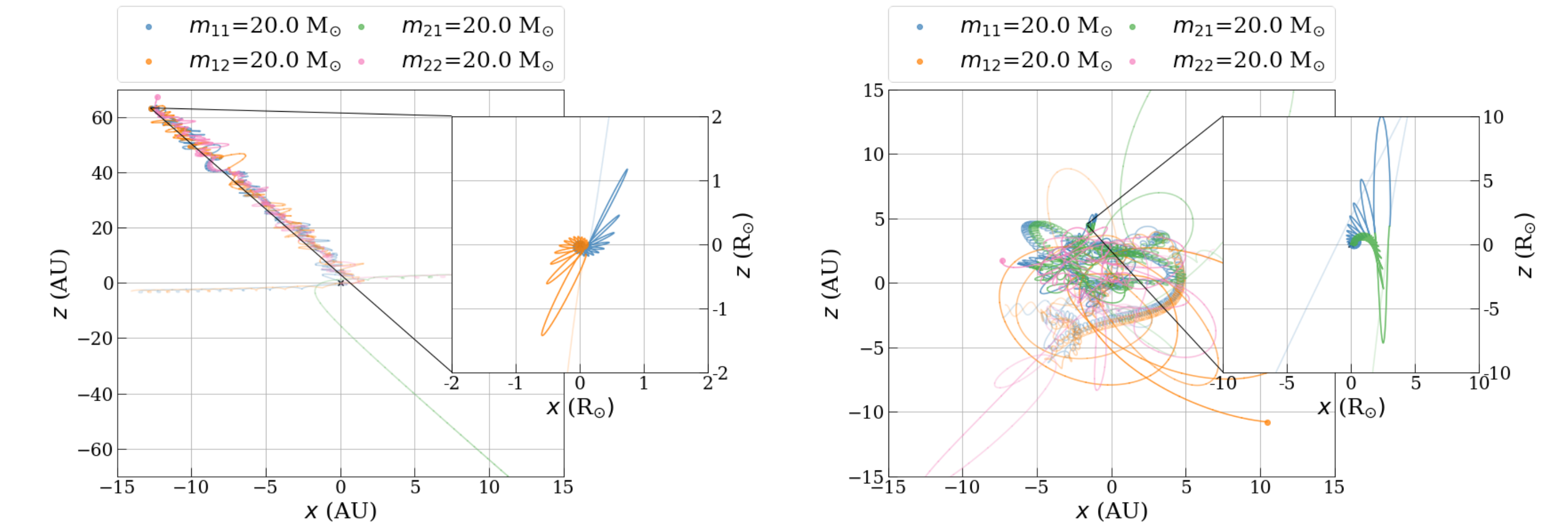}
\caption{Example evolution of binary-binary BH encounters that lead to a GW capture and inspiral. All component masses are 20 $M_{\odot}$, initial binary SMAs are 1\,au, and $v_{\rm \infty}/v_{\rm crit} = 0.01$. Insets zoom into the highly eccentric inspiral. Trajectories are shaded to indicate the passage of time; darker shades correspond to later in the resonant interaction. The encounter on the left takes place over approximately 14\,yr and the encounter on the right over approximately 25\,yr. Animations for these interactions, as well as for other binary-binary interactions from this study, can be found at \url{https://michaelzevin.github.io/media/bbh_progenitors/}. }
\label{fig:trajectories}
\end{figure*}

\subsection{Post-Newtonian Equations of Motion}

The encounters we focus on in this study lead to relativistic velocities and interactions in the strong-field gravity regime; purely Newtonian dynamics fails to capture their true evolution. 
Though there is no general analytic expression for the equations of motion of relativistic bodies, pN theory approximates relativistic effects by formulating the equations of motion in increasing orders of $(v/c)^{\gamma}$. 
The 2.5pN order, which includes terms in the pN expansion with $\gamma=5$, is the lowest pN order at which the dissipative energy effects of GW emission are introduced. 
Prior studies that focused on binary-single BH encounters found that the inclusion of GW emission can lead to GW captures and rapid inspirals \cite[e.g.,][]{Samsing2014}. 

In this study, we include pN terms in the equations of motion up to and including the 2.5pN term \citep[e.g.,][]{Blanchet2014}. 
Though the 2.5pN term is the primary driver facilitating rapid and eccentric mergers during RIs, the 1pN and 2pN terms, which govern periapse precession, may play an important role in the evolution of strong-field four-body encounters. 
Furthermore, precession of the orbit can suppress Lidov-Kozai oscillations in hierarchical triples that lead to mergers with measurable eccentricities \citep{Blaes2002}. 
To ensure the correct implementation of pN terms, we evolve a single BBH system to verify that the evolution of SMA, eccentricity, and angle of periapse match analytical expectations \citep{Peters1964}, and find that the orbit-averaged pN energy is conserved when only pN terms below the 2.5 order are included in the equations of motion \citep{Mora2004}. 
Example binary-binary encounters that led to a GW capture and highly eccentric inspiral are shown in Figure \ref{fig:trajectories}.

\section{Scattering Experiments}\label{sec:scattering_experiments}

With our endstates defined and pN equations of motions implemented, we performed $\mathcal{O}(10^5)$ scatterings for each initial condition, specified by component masses, incoming velocity, orbital SMA, and orbital eccentricity. 
We Monte Carlo sample over all other extrinsic parameters defining the initial configuration of the system (see Section \ref{subsec:initial_conditions}), and accumulate statistics on various endstates and the orbital characteristics of resultant binaries. 

First, we derive analytical approximations and perform scattering experiments to investigate the post-encounter orbital properties from binary-binary interactions. 
Following this, we perform binary-binary scatterings in the strong encounter regime on a fixed grid, varying only one parameter of the system configuration at a time. 
We then consider binary-binary and binary-single encounters from the classical channel of dynamical BBH formation: BBHs assembled in old, metal-poor GCs. 
We use a few dozen GC models with various initial conditions that are evolved over cosmic time using the \verb|Cluster Monte Carlo| (CMC) code \citep{Joshi2000, Chatterjee2010, Morscher2013, Rodriguez2016a}. 
Models are taken from \cite{Rodriguez2018a} and \cite{Rodriguez2018c}, with updates to include orbital dissipation from GWs and 2.5pN terms when integrating strong encounters \citep{Rodriguez2018b}. 
The initial conditions of encounters used in our scattering experiments, as well as the relative abundance of binary-binary and binary-single interactions, are taken from these models. 
We analyze interactions from each cluster model separately, to examine general trends in encounters as a function of cluster property.

\subsection{Orbital Properties Following Strong Encounters}\label{subsec:orbital_properties}

The BHs residing in the collapsed cores of GCs are susceptible to many strong encounters during their lifetimes, thereby erasing information about their primordial orbital histories. 
It is through these strong encounters that binary orbits tighten, as components ejected from the interaction siphon orbital energy during their ionization. 
The distribution of binary orbital properties resulting from strong encounters is thereby largely independent of the properties of primordial binaries, but influenced by the initial energetics of the systems that take part in the strong encounter. 

In the case of strong binary-single interactions, the average change in SMA between the incoming and outgoing binary can be analytically approximated using three-body energetics. 
From the normalized orbital energy distribution for binary systems that are assembled in three-body processes \citep{Heggie1975}, one finds that the mean fractional decrease in binary SMA from a strong binary-single encounter is $\langle \delta_3 \rangle \approx 7/9$ \citep{Samsing2018d}.

\begin{figure}[t!]
\includegraphics[width=0.48\textwidth]{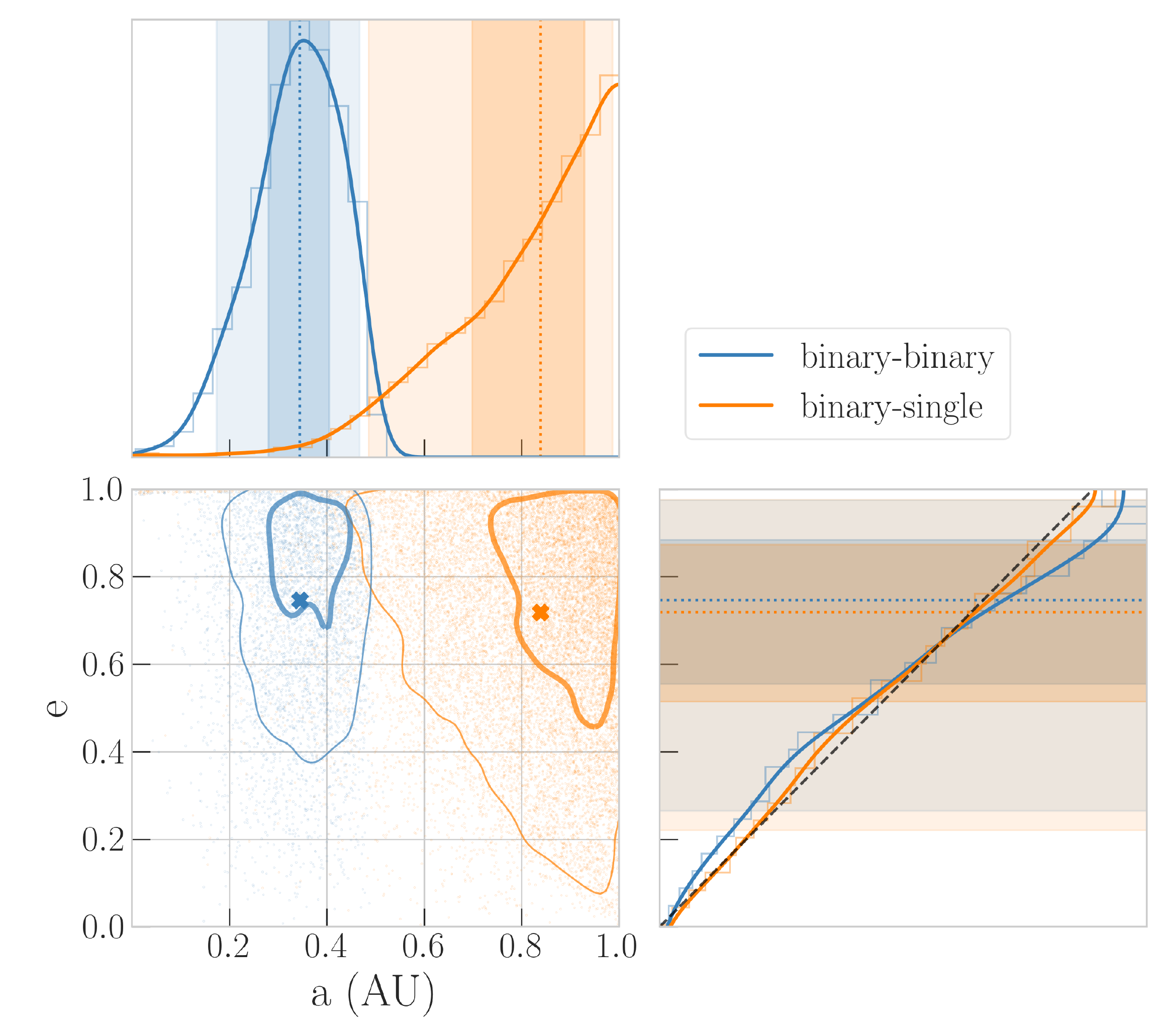}
\caption{Post-encounter orbital properties for bound binaries following binary-binary (blue) and binary-single (orange) exchange and ionization encounters, for fiducial interactions of systems initially on circular orbits with 20 $M_{\odot}$ component BHs, 1\,au initial orbital separation, and $v_{\infty}/v_{\rm crit} = 0.01$. 
In the joint (marginal) distribution, the marker (dotted line), thick line (dark band), and thin line (light band) represent the median, 50\%, and 90\% credible regions. 
A thermal distribution is plotted over the marginal eccentricity distribution with a dashed black line. 
The typical post-encounter SMA for strong binary-binary interactions is clearly separable from the post-encounter SMA of binary-single interactions. }
\label{fig:orbital_parameters}
\end{figure}

In the four-body problem, strong encounters that undergo RIs typically lead to two ejected and unbound components, requiring roughly double the orbital energy to be transferred to the kinetic energy of the ionized particles. 
Furthermore, the greater number of component BHs and larger mass in the interaction means the system's gravitational potential is slightly deeper, and ionization necessitates larger energy requirements. 
Therefore, one would expect the mean fractional decrease in SMA from a strong binary-binary encounter $\left( \langle \delta_4 \rangle \right)$ to be smaller, making the hardening process more drastic in binary-binary encounters. 

Given two incoming binary BHs in the hard binary limit ($v_{\infty} \ll v_{\rm crit}$), the initial energy of the system is determined by the binding energy of the two binaries. 
Assuming equal masses and SMAs, this is given by 
\begin{equation}
E_{0} = -\frac{G m^{2}}{a_{0}}.
\end{equation}
If one of the two binaries becomes dissociated from the encounter and its components ejected, the maximum binding energy that the remaining binary could have is half of the initial energy of the encounter, and therefore
\begin{equation}
\delta_{4, {\rm max}} = 0.5.
\end{equation}

However, the ionized particles are typically ejected at speeds greater than the escape velocity. 
We estimate the typical ejection velocity from a strong encounter through conservation of energy in a binary-single encounter: 
\begin{equation}
-\frac{Gm^2}{2a_0} = -\frac{Gm^2}{2 \langle \delta_3 \rangle a_0} + \frac{1}{2} \mu_3 v_{\rm ion}^2
\end{equation}
where $\mu_3$ is the reduced mass of the three-body system ($2m/3$ for equal mass) and $v_{\rm ion}$ is the typical velocity of the ionized particle. 
Solving for $v_{\rm ion}$ yields
\begin{equation}\label{eq:Vion}
v_{\rm ion} = \sqrt{\frac{3 G m}{2 \langle \delta_3 \rangle a_0}(1-\langle \delta_3 \rangle)}. 
\end{equation}

If we assume that the velocity of the ionized particles after the encounter is similar for binary-binary interactions, equating initial and final energy of an ionizing binary-bianry interaction gives us 
\begin{equation}
-\frac{Gm^2}{a_0} = -\frac{Gm^2}{2 \langle \delta_4 \rangle a_0} + \frac{1}{2} (\mu_3 + \mu_4) v_{\rm ion}^2
\end{equation}
where $\mu_4$ is the reduced mass of the three-body metastable system and the ionized particle ($3m/4$ for equal mass). 
Substituting $v_{\rm ion}$ from Equation \ref{eq:Vion} and solving for $\langle \delta_4 \rangle$, we get 
\begin{equation}
\langle \delta_4 \rangle = \frac{24 \langle \delta_3 \rangle}{51 - 3 \langle \delta_3 \rangle}. 
\end{equation}

If we assume a binary-single encounter does not harden the resultant binary at all, $\langle \delta_3 \rangle$ = 1 and $\langle \delta_4 \rangle$ reduces its maximum value of $\delta_{4, {\rm max}} = 0.5$. 
Using instead a value of $\langle \delta_3 \rangle \approx 7/9$ as derived in \cite{Samsing2018d}, we find a value of  $\langle \delta_4 \rangle \approx 0.38$ where again $\langle \delta_4 \rangle$ is the mean fractional change in the SMA of the remaining binary after an ionizing four-body encounter. 
Alternatively, one could instead assume the ejection process results from two successive ejections by one of the binaries in the interaction, which each have an equal effect on the hardening process. 
In this case, $\langle \delta_4 \rangle$ is simply found to be $0.5 \times \langle \delta_3 \rangle^2 = 0.30$. 

These analytical approximation are supported by the scattering experiments shown in Figure \ref{fig:orbital_parameters}. 
Here, we plot the post-interaction orbital properties of bound binaries that underwent a binary-binary (blue) and binary-single (orange) exchange and ionization.\footnote{We focus on the exchange and ionization endstate because this guarantees that the binary-single system went through a strong encounter. Since we only track the resultant particle configurations, if the output configuration is identical to the input configuration, it is ambiguous whether a resonating encounter and ionization occurred, or simply a weak fly-by.} 
As expected, the bound post-interaction binary eccentricities follow a thermal distribution. 
For the post-interaction SMAs, we find median and 90\% credible values of $0.84  ^{+0.12}_{-0.21}$ and $0.34 ^{+0.09}_{-0.10}$ for $\langle \delta_3 \rangle$ and $\langle \delta_4 \rangle$, respectively, consistent with our analytical approximations. 
Although our simple estimates seem to provide a good understanding of the expected hardening of the remaining binary after a strong binary-binary encounter, we refer the reader to \cite{Leigh2016} and \cite{Leigh2017} for a more detailed study of the problem.

\subsection{Fiducial Strong Encounters}\label{subsec:fiducial_strong_encounters}

In the following two sections, we explore how the probability of inducing a GW captures is affected by the initial conditions of the interacting system using fiducial binary parameters, and quantitatively examine how non-dissipative pN terms influence the cross section of such GW captures.

\subsubsection{Dependence on Binary Parameters}

\begin{figure}[t!]
\centering
\includegraphics[width=0.5\textwidth]{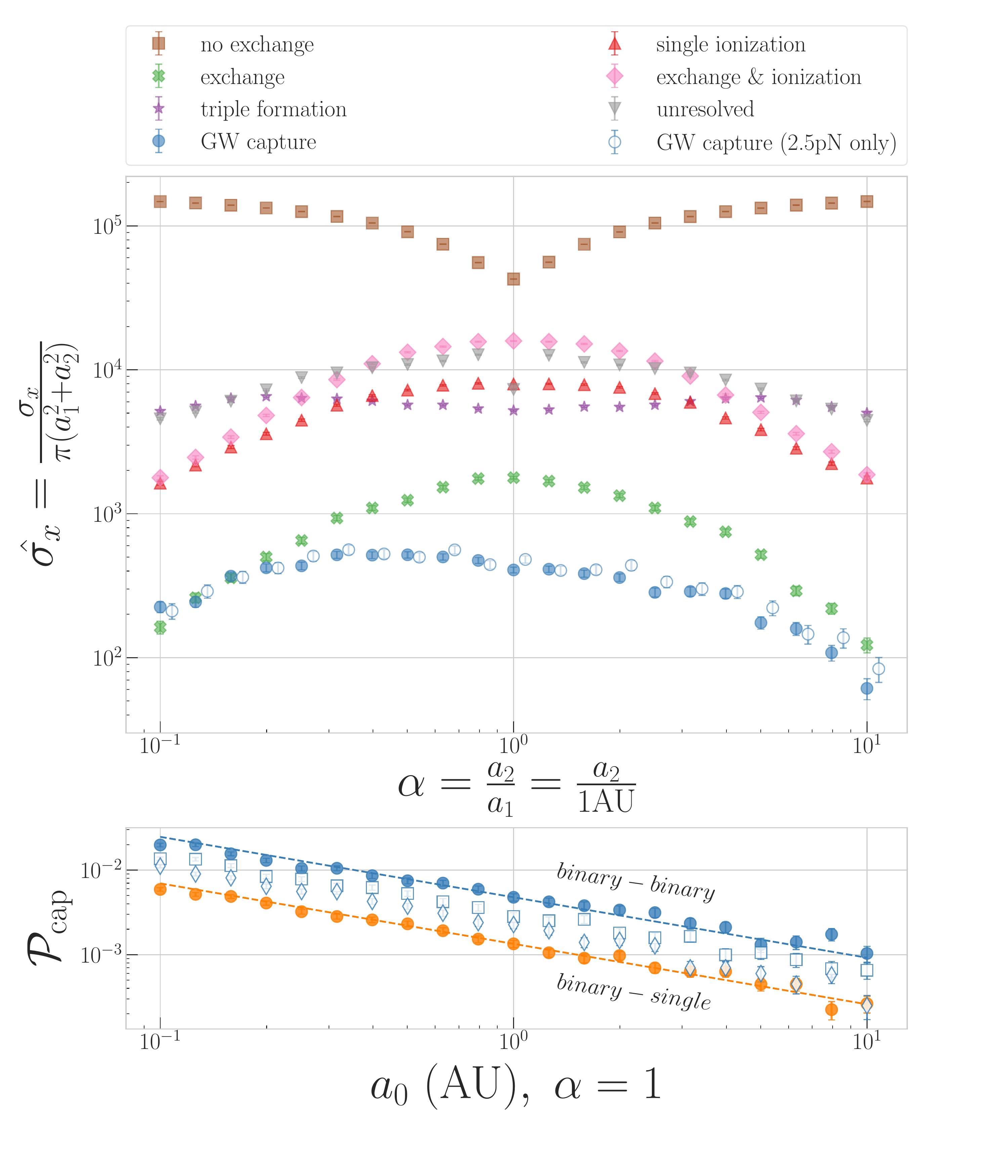}
\caption{Reduced cross sections for binary-binary endstates as a function of SMA ratio $\alpha$ (top panel) and GW capture probability for binary-binary and binary-single encounters as a function of SMA (bottom panel) in the hard binary limit. 
All systems have component masses of $m$ = 20 $M_\odot$ and incoming velocities of $v_{\infty}/v_{\rm crit} = 0.01$, with impact parameters sampled according to Equation \ref{eq:inpact_parameter}. 
Because reduced cross sections are normalized by the orbital area and $v_{\infty}/v_{\rm crit}$ is fixed, the scaling relations are identical to those for a particular endstate probability.
Top panel: the target binary has a fixed SMA of $a_{1}=1$\,au. 
We find that the single ionization and exchange endstates scale as $\alpha^{-1}$ and $\alpha^{-2}$, respectively, as found in \cite{Antognini2016}. 
For the GW capture endstate, we show cross sections for simulations where we include all pN terms up through the 2.5pN order (filled circles) and only the 2.5pN order (open circles, artificially offset for readability); we find no statistically significant change due to the inclusion of lower-order non-dissipative pN terms. 
Bottom panel: SMAs are varied between 10$^{-1}$ and 10$^{1}$\,au, with an SMA ratio of unity. 
The reduced cross sections of the GW capture endstate scales as $a^{-5/7}$, analogous to binary-single encounters \citep[cf.][]{Samsing2014}. 
Also shown in open squares and open triangles are binary-binary GW capture probabilities where the component binary mass ratios are 0.5 and 0.25, respectively, and the total mass of each binary in the interaction is held fixed at 40 $M_\odot$. 
These systems follow the same scaling relation as the equal mass case, but with slightly lower GW capture probability. 
}
\label{fig:cross-sections}
\end{figure}

Figure \ref{fig:cross-sections} shows the scaling of endstate cross sections as a function of SMA and SMA ratio, $\alpha=a_2/a_1$. 
Though incoming velocities also affect endstate probability, the velocity dispersion within the cores of GCs are typically low compared to the critical velocity of the binaries. 
To this end, we examine a grid of binaries with equal mass ratios in the hard binary limit ($v_{\infty} \ll v_{\rm crit}$), with fiducial values of $v_{\infty}/v_{\rm crit} = 0.01$ and component BH masses of $m_{ij}$ = 20 $M_{\odot}$.

We find the expected scaling relations derived for binary-binary encounters in the Newtonian regime (cf. Figure 3 in \cite{Antognini2016}).  
The GW capture endstate, which was not included in Newtonian scattering experiments, reaches a peak probability at $\alpha \lesssim 1$ and occurs at a probability approximately two orders of magnitude less than the most probable endstate at equal SMA ratio (no exchange). 
For values of $\alpha \gg 1$ or $\alpha \ll 1$, we find that the GW capture cross section once again drops; this is due to the tighter binary effectively acting as a single particle during the interaction, and therefore the encounter proceeds similarly to a three-body interaction. 
This will cause shorter-lived RIs with less IMSs than a typical four-body encounter, thereby decreasing the probability of the GW capture endstate. 

Notably, the GW capture endstate occurs at a higher probability for values of $\alpha < 1$ compared to large values of $\alpha$, whereas the Newtonian endstates are all symmetric about $\alpha$. 
The bottom panel of Figure \ref{fig:cross-sections} shows the reason for this effect. 
We define the GW capture probability simply as
\begin{equation}\label{eq:inspiral_probability}
\mathcal{P}_{{\rm cap}, j} = \frac{N_{{\rm cap}, j}}{N_{{\rm total}, j}}
\end{equation}
where $N_{{\rm total}, j}$ is the total number of either binary-binary ($j$=bb) or binary-single ($j$=bs) scattering experiments that are performed and $N_{{\rm cap}, j}$ is the number of those encounters that lead to a GW capture endstate. 
Thus, the reduced cross section for an inspiral is given by 
\begin{equation}
\hat{\sigma}_{{\rm cap}, j} = \mathcal{P}_{{\rm cap}, j} \hat{\sigma}_{\rm CI} \simeq \mathcal{P}_{{\rm cap}, j} \frac{3 G M}{a_{0} v_{\rm \infty}^{2}}
\end{equation}
where $M$ is the total mass of the four-body system, $a_{0}$ is the initial SMA of the target binary, and $\sigma_{\rm CI}$ is the cross section for a close interaction where a system passes within a sphere of influence marked by the target binary's separation \citep{Samsing2014}.

For an IMS binary to undergo a GW capture during the RI, its pericenter distance must be below some characteristic capture distance $r_{\rm cap}$.
The value of this distance is determined from where the GW energy loss integrated over one pericenter passage, $\Delta{E}_{\rm p}(r_{\rm p}) \approx (85\pi/12)G^{7/2}c^{-5}m^{9/2}r_{\rm p}^{-7/2}$ \citep[see][]{Hansen1972}, is comparable to the total energy of the few-body system \citep{Samsing2017e, Samsing2017d}, which in the hard binary limit is approximately the binding energy of the initial binaries, $E_{\rm B}(a_0) \propto m^{2}/a_0$ \citep[see][]{Samsing2014}. 
Solving for the pericenter distance at which $\Delta{E}_{\rm p}(r_{\rm cap}) = E_{\rm B}(a_0)$, one now finds $r_{\rm cap} \propto m^{5/7} a_0^{2/7}$ \citep{Samsing2017d}. 
Because the eccentricity of the IMS binary follows a thermal distribution, the probability for a GW capture and inspiral is $\mathcal{P}_{\rm cap} \propto r_{\rm cap}/a_0$. 
Thus, the reduced cross section for a GW capture endstate scales as 
\begin{equation}
\hat{\sigma}_{{\rm cap}, j}=\mathcal{P}_{{\rm cap}, j}\hat{\sigma}_{\rm CI} \propto \frac{m^{12/7}}{a_0^{12/7} v_{\infty}^2} \propto \left(\frac{m}{a_0}\right)^{5/7}
\end{equation}
where the last proportionality holds for our scattering experiments where the incoming velocity is fixed relative to the critical velocity of the target binary, which scales as $v_{\rm crit} \propto \sqrt{m/a_0}$. 
Thus, for fixed $v_{\infty}/v_{\rm crit}$ the reduced cross section is directly proportional to the GW capture probability.

This behavior is consistent between binary-binary and binary-single encounters, as can be seen in the bottom panel of Figure \ref{fig:cross-sections}. 
The peak in the GW capture probability for values of $0.3 \lesssim \alpha \lesssim 0.8$ is due to the interplay of the two effects described above: though the GW capture probability increases as the SMA decreases, if $\alpha$ becomes too small then the tighter binary will act as a single particle and the interaction will effectively proceed as a three-body encounter. 
Most important, though, is that we find the probability of a binary-binary encounter leading to a GW capture is 3--5 times higher than the probability of a binary-single encounter leading to a GW capture for any initial value of the SMA. 

To test the sensitivity of the GW capture cross-section scaling relation to variations in binary properties, scatterings with unequal masses and SMAs were also examined. 
Though Figure \ref{fig:cross-sections} only shows binaries with an SMA ratio of $\alpha=1$, SMA ratios of $\alpha=0.1$ and $\alpha=0.5$ found consistent GW capture probabilities and an identical scaling relation --- the scaling of the GW capture probability as a function of SMA in the bottom plot is therefore true for each SMA ratio in the top plot.
Moving to mass ratios further away from unity while keeping the total mass fixed causes the GW capture probability to decrease; though the slope of the power law remains the same ($\mathcal{P}_{\rm cap} \propto a_0^{-5/7}$), the GW capture probability is $\sim$2--3 times lower when the mass ratio drops to 0.25. 
This is likely due to the lower-mass components being preferentially ejected during the interaction, leading to a shorter-lived RI and fewer IMS binaries synthesized during the interaction, and therefore a slimmer chance of drawing the high eccentricity needed for a rapid GW capture before the system dissociates \citep{Sigurdsson1993}. 
However, binaries that interact in cluster cores trend toward equal masses, so we expect the scatterings with a mass ratio of unity to be most representative of the true cluster binaries. 
In fact, in our cluster models, we find a median mass ratio of $\sim$0.9 in the component binaries of binary-binary interactions, with 90\% of systems having mass ratios greater than $\sim$0.6.

\pagebreak
\subsubsection{Effect of Non-Dissipative pN Terms}\label{subsec:nondissipative}

We also examine how the inclusion of the 1pN and 2pN terms in the equations of motion affect the induction of GW captures during RIs. 
The reduced cross section for the GW capture endstate including (not including) the 1pN and 2pN terms can be seen in Figure \ref{fig:cross-sections} with filled (unfilled) blue circles. 
We find no measurable difference in the GW capture cross section when 1pN and 2pN terms are included, compared to simulations where only the 2.5pN term is included. 
Though the statistical and systematic uncertainty affect our measurement of the GW capture cross section to a higher degree, our experiments show that the amplification or suppression of GW captures during RIs due to the inclusion of non-dissipative pN terms is limited to percent-level deviations at most. 
However, it is important to note that the 1pN and 2pN terms are crucial for following the evolution of encounters that result in stable triples. 
Though we do not examine the evolution of such systems here, we refer the reader to \cite{Antonini2016a}. 

\subsection{Strong Encounters in GCs}\label{subsec:GC_encounters}

We next perform binary-single and binary-binary BH scattering experiments tuned to GC models, which provide a distribution of pre-encounter orbital parameters that is more representative of systems in the universe. 
In particular, we gauge the relative contributions of binary-binary and binary-single encounters in inducing GW captures. 
Regardless of cluster properties, binary-single BH encounters occur $\sim$10--100 times more frequently than binary-binary BH encounters. 
However, as seen in Figure \ref{fig:cross-sections}, the probability of GW captures is higher for the given binary-binary encounters by a factor of $\sim$3--5. 
For our scattering experiments involving cluster binaries, we only include the 2.5pN term, as the orbital characteristics are more accurately extracted, and as shown in Section \ref{subsec:nondissipative}, non-dissipative pN terms have a negligible effect on the cross section of GW inspirals during RIs.

\begin{figure}[t!]
\includegraphics[width=0.48\textwidth]{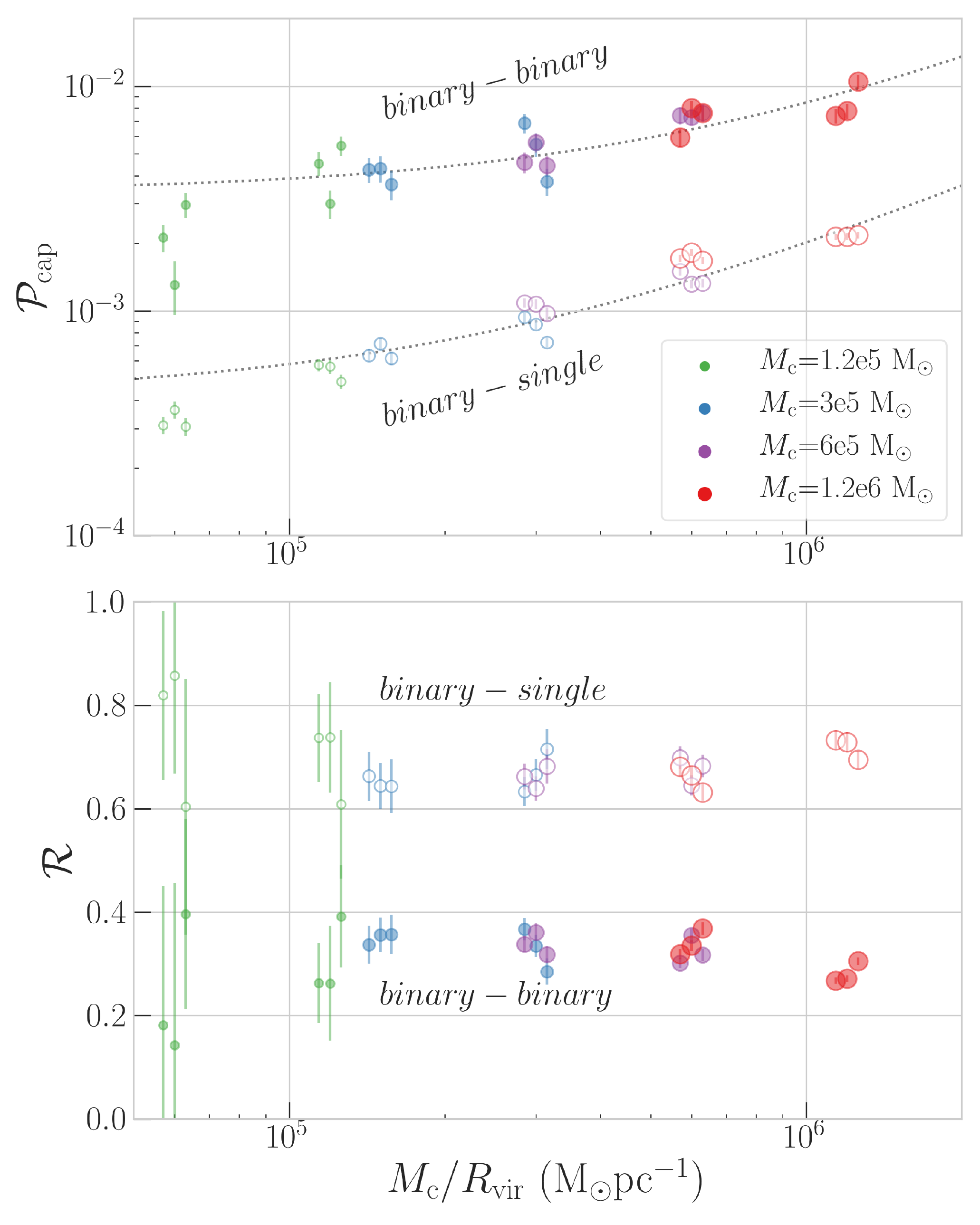}
\caption{GW capture probability (top panel, Equation \ref{eq:inspiral_probability}) and GW capture contribution (bottom panel, Equation \ref{eq:inspiral_contribution}) for binary-binary (filled circles) and binary-single (open circles) encounters in GCs with varying masses, virial radii, and metallicities. 
Cluster models with different masses are differentiated by colored circles of different sizes. 
For each cluster compactness, we artificially offset clusters with different metallicity for readability; for each compactness value, the leftmost point is for a metallicity of $Z=0.0005$, the middle point for $Z=0.001$, and the right point for $Z=0.005$. 
Dotted lines in the top plot show linear fits to the data. 
We take into account statistical uncertainty in both $\mathcal{P}_{{\rm cap},j}^{\rm c}$ and $N_{j}^{\rm c}$. }
\label{fig:efficiency}
\end{figure}

\subsubsection{Inspiral Efficiency of Binary-Binary Interactions}

In the top panel of Figure \ref{fig:efficiency}, we show the GW capture probability for binary-binary and binary-single encounters in a range of cluster models with various masses, metallicities, and virial radii. 
The cluster compactness is the dominant influence on GW capture probability, with metallicity playing no noticeable role; we therefore plot $\mathcal{P}_{\rm cap}$ for the various cluster models as a function of compactness, defined as $M_{\rm c}/R_{\rm vir}$, where $M_{\rm c}$ is the initial GC mass and $R_{\rm vir}$ its initial virial radius. 
Similar to our gridded scattering experiments, we find that, in our GC models, the probability of inducing a GW capture during a binary-binary interaction is larger than that for a binary-single interaction. 
In fact, from our strong encounters extracted from GC models, which are more representative of the true astrophysical population, the value is found to be even higher; these experiments indicate that a binary-binary interaction is $\sim$4--7 times more likely to lead to a GW capture than a binary-single encounter. 

We can define the contribution of binary-single or binary-binary GW captures to the total number of GW captures in a given cluster model as 
\begin{equation}\label{eq:inspiral_contribution}
\mathcal{R}_{j}^{\rm c} = \frac{\mathcal{P}^{\rm c}_{{\rm cap},j} N^{\rm c}_{j}}{\sum\limits_{k} \mathcal{P}^{\rm c}_{{\rm cap},k} N^{\rm c}_{k}}
\end{equation}
where $\mathcal{P}^{\rm c}_{\rm cap,j}$ is the GW capture probability and $N^{\rm c}_{\rm j}$ is the total number of binary-single ($j$=bs) or binary-binary ($j$=bb) interactions in a given cluster model. 
We find binary-binary GW capture contributions of $\mathcal{R}_{\rm bb}^{\rm c} \sim $25--45\% in the GC models examined, with a median of 36\%. 
Though we find $\mathcal{R}_{\rm bb}^{\rm c}$ to slightly increase as $M_{\rm c}/R_{\rm vir}$ decreases, the general properties of the GC environment have little effect on $\mathcal{R}_{\rm bb}^{\rm c}$. 

Analytical arguments in \cite{Samsing2018d} predict that the GW capture probability scales linearly with the cluster compactness. 
As seen in Figure \ref{fig:efficiency}, this linear trend fits our data well for GC models with moderate and high compactness, though the GW capture probability is lower than expected for clusters with low compactness. 
This deviation may be due to small number statistics, as fewer strong interactions take place during the lifetime of a low-mass cluster.

\subsubsection{Eccentricity of Inspirals}\label{subsec:inspiral_populations}

One of the most notable properties of GW inspirals that arise from dynamical encounters is their eccentricity. 
We divide the BBH mergers from such encounters into three categories:

\begin{itemize}

\item
\textsc{Ejected Inspirals} are binary systems whose post-encounter center of mass velocity exceeds the escape velocity of the GC,\footnote{The escape velocity of a GC can change drastically as the cluster evolves over cosmic time. For each interaction in the GC models, the escape velocity from the location of the interaction at the time of the interaction is recorded, and this is compared to the post-encounter center of mass velocity to determine whether the system is ejected from the cluster. } and they evolve in isolation following their ejection. 
We only include systems that merge within a Hubble time in this population. 
As these systems merge over timescales ranging from tens of millions to billions of years, they have mostly circularized by the time GW emission evolves their orbits into the LIGO/Virgo sensitive frequency range, regardless of their post-encounter eccentricity. 

\item
\textsc{In-Cluster Binary Inspirals} leave a resonating encounter in a hardened binary system with post-encounter velocities that do not exceed the escape speed of the cluster. 
If the SMA is small enough and/or the eccentricity is large enough, these binaries can merge through GW emission before encountering another object in the cluster. 
The typical interaction timescale of objects in the core of a cluster is dependent on cluster properties and the age of the cluster, and the precise time between discrete interactions was not recorded in our models. 
However, we can approximate the typical interaction timescale of a binary from its cross section, which is dependent on the component masses and SMA of the binary. 
Taking a fiducial interaction timescale of $\tau = 10$\,Myr for a post-encounter binary with $m_1$ = $m_2$ = 20 $M_{\odot}$ and $a = 0.3$\,au \citep[see][]{Samsing2018d}, and noting that the binary-single interaction cross section is given by $\sigma_{\rm bs} = 6 \pi G m a / v_{\rm disp}^2$ where $v_{\rm disp}$ is the velocity dispersion, the typical time between binary-single interactions scales as 
\begin{equation}
\tau_{\rm int} \approx \frac{1}{n_{\rm BH} \sigma_{\rm bs} v_{\rm disp}} \propto \frac{v_{\rm disp}}{n_{\rm BH} m a}
\end{equation}
where $n_{\rm BH}$ is the number density of single BHs. 
We scale the fiducial value of $\tau_{\rm int}$ by the total mass and SMA accordingly for each post-encounter binary and compare it to the GW merger timescale: 
\begin{flalign}\label{eq:inspiral_time}
\tau_{\rm insp}&(a_0,e_0) = \\ 
&\frac{12}{19} \frac{c_0^4}{\beta} \int_{0}^{e_0} \frac{e^{29/19}[1+(121/304)e^2]^{1181/2299}}{(1-e^2)^{3/2}}de,\nonumber
\end{flalign}
where $\beta$ is a constant factor dependent on the component masses and $c_0$ is determined by the initial conditions $a=a_0$ and $e=e_0$ \citep{Peters1964}. 
If $\tau_{\rm insp} < \tau_{\rm int}$, the binary merges prior to its next encounter in the cluster and is therefore classified as an in-cluster binary inspiral. 
Because these tight binaries merge on a shorter timescale than ejected binaries, they typically merge with larger eccentricities than those of ejected mergers.
\item

\textsc{GW Captures}\footnote{In recent work, this category of GW inspirals are often referred to as ``three-body mergers.'' However, because this work shows that a significant fraction of such mergers come from encounters involving more than three bodies, we instead use the nomenclature ``GW captures'' for these systems.} are systems that inspiral and merge during the RI itself, determined when a GW capture endstate is reached (see Section \ref{subsec:endstates}), or if a binary inspirals and merges within 10$^5$ s of its final encounter in the RI.\footnote{We include this criterion because the final interaction of the RI, which would eventually leave the binary in isolation, sometimes induces the highly eccentric inspiral that mergers shortly after the isolation tidal threshold is reached and the system is marked as an ionization endstate. Without this criterion implemented, we would find a small number of highly eccentric systems that are categorized as in-cluster binary inspirals rather than GW captures. }
This can occur from the formation of a hard eccentric IMS binary that merges during the chaotic encounter, or through a highly eccentric capture where the two objects emit enough gravitational radiation during a close pass on a hyperbolic orbit for the binary to become bound and rapidly inspiral. 
This mechanism can even cause highly eccentric binaries to be formed \textit{within} the sensitive frequency ranges of ground-based GW detectors, resulting in initial eccentricities close to unity and mergers that typically occur less than a second after the system becomes bound. 
\end{itemize}

\begin{figure*}[t!]
\includegraphics[width=1.0\textwidth]{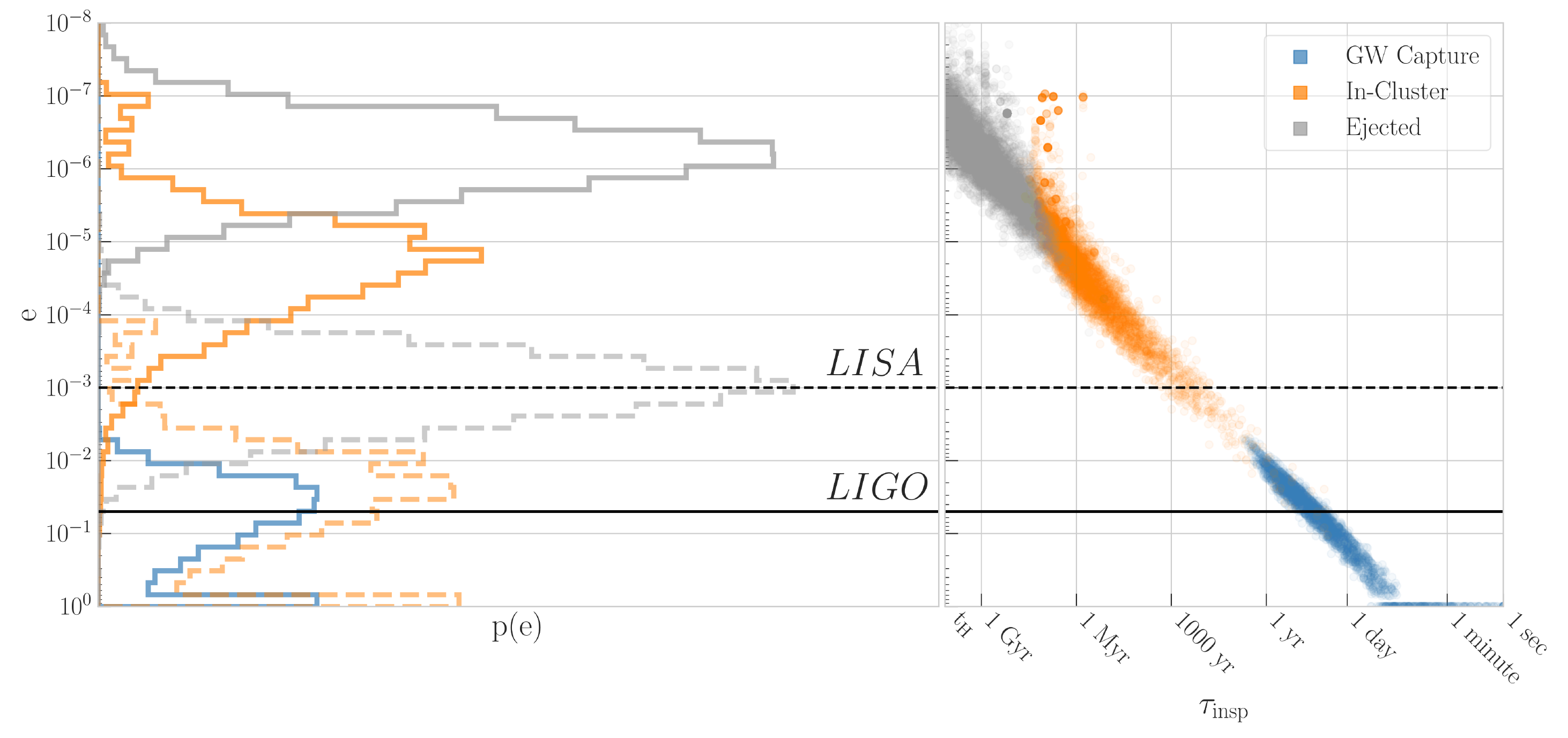}
\caption{Eccentricity distributions and delay times resultant from three distinct populations of binary-binary induced GW inspirals: ejected inspirals (gray), in-cluster binary inspirals (orange), and GW captures (blue). Solid colored lines show the eccentricity at 10 Hz --- the approximate lower end of the LIGO/Virgo sensitive frequency range. Dashed colored lines show the eccentricity at 10$^{-2}$ Hz --- the most sensitive frequency predicted for the space-based LISA detector. The solid and dashed black lines show minimum measurable eccentricities predicted for LIGO/Virgo \citep{Lower2018} and LISA \citep{Nishizawa2016}, respectively. None of the GW captures are accessible by LISA, as they form at orbital frequencies above LISA's sensitive frequency range. }
\label{fig:eccentricity}
\end{figure*}

We record the eccentricities and SMAs of each bound binary once an endstate is reached in our scattering simulations. 
To find the eccentricity at a particular GW frequency, we numerically solve for the orbital properties of the binary at a given eccentric peak frequency as in \cite{Wen2003}: 
\begin{equation}
a(e) = \frac{1}{1-e^{2}}\left[\frac{G M}{\pi} \frac{(1+e)^{1.1954}}{f_{\rm GW}}\right]^{2/3},
\end{equation}
which is coupled to the differential equation from \cite{Peters1964} 
\begin{equation}
\left\langle \frac{da}{de} \right\rangle = \frac{12}{19} \frac{a}{e} \frac{[1+(73/24)e^2 + (37/96)e^4]}{(1-e^2) [1+(121/304)e^2]}.
\end{equation}

For cases where the binary forms at frequencies above $f_{\rm GW}$, $e(f_{\rm GW}) > 1$ and these equations are not differentiable.
To distinguish these systems, we calculate the pericenter distance at a reference eccentricity of $(1-e_{\rm ref})=10^{-3}$. The pericenter distance is given by \cite{Peters1964}: 
\begin{equation}
R_{\rm p} = (1-e) a(e) = (1-e)\frac{c_0 e^{12/19}}{(1-e^2)} \left[ 1+\frac{121}{304}e^2 \right]^{870/2299}
\end{equation}
where the constant $c_0$ is determined by the orbital parameters recorded at the end of the integration. 
We compare this distance to the SMA of the binary if it were on a circular orbit at an orbital frequency of $f_{\rm orb} = f_{\rm GW}/2$: 
\begin{equation}
a_{\rm circ} = \frac{G M}{\pi^2 f_{\rm GW}^2}.
\end{equation}
where $f_{\rm GW}$ is the minimum detectable frequency by a fiducial GW detector. 
If $R_{\rm p} < a_{\rm circ}$, then the binary formed inside the sensitive frequency range of the detector at a frequency greater than $f_{\rm GW}$, and it is assigned an extremal eccentricity of $e=1$. 
Because the $R_{\rm p}$ asymptotically approaches the true initial pericenter distance as $e$ approaches 1, our choice of reference eccentricity has little effect on this procedure. 
For example, increasing or decreasing $(1-e_{\rm ref})$ by an order of magnitude changes the value of $R_{\rm p}$ by less than 0.5\%.

Similar to \cite{Samsing2017c} and \cite{Rodriguez2018b}, we find GW inspirals induced from pN binary-binary RIs to have a distinctive imprint on the distribution of binary eccentricities. 
Figure \ref{fig:eccentricity} shows the eccentricity distributions for these three categories of dynamically induced inspirals from binary-binary encounters. 
We find little difference in the shape of the binary-binary and binary-single eccentricity distributions in our simulations, and the highly eccentric peak in our binary-binary distributions is consistent with previous work \citep[][]{Samsing2017c,Rodriguez2018b}.

We also calculate the inspiral time ($\tau_{\rm insp}$) from the formation of the binary until the merger as in Equation \ref{eq:inspiral_time}. 
However, because we record the binary information at the point when the simulations terminate, the orbital properties at binary formation for systems that reached a GW capture endstate are ambiguous. 
To estimate the inspiral times of this population, we note that these systems are generally formed from highly eccentric captures, and therefore choose a high reference eccentricity of $(1-e_{\rm ref})=10^{-3}$ at formation and calculate $a_{\rm ref}$ and $\tau_{\rm insp}$ accordingly.

The inspiral times for our three populations of GW inspirals can be seen in the right panel of Figure \ref{fig:eccentricity}. 
The continuum of inspiral times between low-eccentricity GW captures and high-eccentricity in-cluster binary inspirals supports our choice for $e_{\rm ref}$. 
Additionally, we find that the inspiral times of GW captures are relatively insensitive to our choice of $e_{\rm ref}$. 
We comment on this further, as well as other methods that have been utilized for calculating the inspiral times of GW captures, in Section \ref{sec:discussion}.

\section{Discussion}\label{sec:discussion}

Until recently, the prospect of detecting eccentric BH mergers with ground-based GW detectors was stifled by assumed long inspiral times and the ensuant damping of orbital eccentricities. 
However, substantial progress has been made over the past couple years in identifying formation scenarios that can maintain appreciable eccentricity even at the high GW frequencies accessible to ground-based detectors, such as the merger of binaries in hierarchical triples from the Lidov-Kozai mechanism \citep{Antonini2016a,Silsbee2017,Randall2018}, single-single captures in NSCs \citep{OLeary2009,Lee2010,Kocsis2012}, and RIs in the cores of stellar clusters \citep{Rodriguez2018b,Samsing2018d}.
These channels all predict astrophysical rates within the predicted detection capabilities of the Advanced LIGO/Virgo network at design sensitivity \citep{Abadie2010}, and may be the most important feature in a GW waveform for definitively distinguishing its BBH progenitor from an isolated binary formation scenario. 
Prior to this study, the role that strong binary-binary encounters play in inducing eccentric GW inspirals was never systematically explored.

\subsection{GW Capture Contribution from Binary-Binary Encounters}

The literature on pN scattering experiments has thus far focused on the contribution of binary-single encounters as the main driver of GW captures in GCs. 
By performing scattering experiments initialized on binary-single and binary-binary encounters extracted from cluster models, we were able to quantify the importance of binary-binary encounters in the pN scenario for the first time. 
Even though binary-binary encounters have a larger interaction cross section, the relative scarcity of binary BH systems compared to isolated BHs in cluster cores lead to these interactions occurring an order of magnitude less often than binary-single BH interactions. 
However, binary-binary interactions lead to more complex RIs that last significantly longer than their binary-single counterparts, facilitating $\sim$\,5 times more metastable IMS binaries before the system dissociates. 
Because the probability of generating a GW capture and inspiral scales linearly with the number of IMSs, this causes binary-binary encounters to be $\sim$\,5 times more likely to induce a highly eccentric inspirals. 
This culminates in a 25--45\% amplification in the rate of GW captures predicted solely from binary-single interactions \citep{Samsing2018d}.
The relative rate of GW captures in GCs is therefore expected to be $\sim$\,10\% with approximately half of these mergers having eccentricities large enough to be measurable by LIGO/Virgo. 
These numbers have also been found using pure numerical techniques \citep{Rodriguez2018b}; the remarkably consistency between these different approaches indicates the robustness of this result. 

From the bottom panel in Figure \ref{fig:efficiency}, we see that the binary-binary contribution is weakly sensitive to the compactness of the cluster environment; binary-binary encounters contribute $\sim$\,40\% of the total number of eccentric inspirals from GW captures for clusters with $M_{\rm c}/R_{\rm vir} \sim 10^5$ compared to $\sim$\,30\% for clusters with $M_{\rm c}/R_{\rm vir} \sim 10^6$. 
As expected, we find metallicity have no effect on the relative importance of binary-binary and binary-single encounters at inducing GW captures.
Despite this moderate variability, the relatively weak sensitivity on the specifics of cluster properties indicates that the binary-binary contribution to GW captures found in our investigation is a good measure of the contribution in the true, astrophysical GC population. 

One interesting property of GCs that may affect the role and rate of binary-binary encounters is the initial stellar binary fraction. 
The GC models in this study use an initial binary fraction of 5\%, which is loosely based on the observed binary fraction of Milky Way GCs \citep{Rubenstein1997,Bellazzini2002,Ivanova2005}. 
Though this may be representative of the initial binary fraction for low-mass stars, it is not necessarily the case for high-mass stars that are the progenitors of BHs. 
The high-mass binary fraction is much more opaque, as progenitor stars of high-mass primordial binaries finished their stellar evolution early in the history of the cluster and are no longer observable in the local universe. 
However, \cite{Chatterjee2016a} found that, regardless of initial high-mass binary fraction, GCs converge to the same number of remaining BBH systems after $\sim$\,3\,Gyr of evolution, further supporting the evidence that BH interactions at late times are set by the internal dynamics and overall cluster properties rather than the details of the initial binary orbital properties. 
Nevertheless, we ran a set of GC models where the primordial high-mass binary fraction was varied between 5\% and 100\%, tracking the number of binary-single and binary-binary BH interactions over cosmic time. 
As expected, we find little difference in both the total number of BH interactions or the relative contribution of binary-binary encounters in facilitating inspirals as we change the initial binary fraction. 

Another way in which binary-binary interactions may lead to more in-cluster mergers is through triple formation. 
In the Newtonian regime, a binary-single encounter will not be able to synthesize a bound, stable triple system. 
Though energy losses through GW emission can theoretically lead to a bound triple from a strong binary-single BH interaction, the presence of an extra component in the interaction makes binary-binary encounters much more efficient at generating bound triple systems \citep{Fregeau2004,Antognini2016}, as the ejection of the fourth BH can efficiently drain energy from the encounter and result in a bound, stable three-body state (see the triple cross section in Figure \ref{fig:cross-sections}). 
If the Lidov-Kozai oscillation timescale is significantly shorter than the periapse-precession timescale, the third body can induce a highly eccentric mergers in the inner binary before the outer binary of the triple is disrupted due to another encounter in the cluster core \citep{Antonini2016a}. 
Mergers from Lidov-Kozai oscillations in triples will likely imprint a unique eccentricity distribution relative to other in-cluster mergers. 
This amplification will be investigated further in future work.

\subsection{Effect of Cluster Properties on Highly Eccentric GW Inspirals}

Though cluster properties do not strongly affect the relative fraction of eccentric inspirals between binary-binary and binary-single BH encounters, the GW capture cross section and probability are sensitive to the cluster property particulars. 
As anticipated in \cite{Samsing2018d}, we find the cluster compactness to be the primary influence on the GW capture probability in a given cluster. 
As cluster cores become more compact, the escape velocity necessary to eject binaries from strong encounters will increase, which leads to binaries achieving harder orbits before it is energetically probable for them to be ejected. 
We find that the GW capture probability for a BH encounter scales linearly for clusters with moderate to high compactness values in both the binary-binary and binary-single cases, though as can be seen in Figure \ref{fig:efficiency}, extrapolating this relationship to low compactness values slightly overpredicts the GW capture probability. 

The GW capture probabilities from binary-binary encounters in the clusters we examine range from $\sim$\,0.002--0.01. 
Therefore, in our most optimistic models, a binary-binary BH encounter will lead to a GW capture and eccentric inspiral approximately once every 100 binary-binary encounters. 
However, massive and compact clusters are highly efficient at ejecting their BHs \citep{Chatterjee2016a}. 
In the local universe, BHs are more likely to reside in the cores of more diffuse clusters, where an appreciable number of BHs may still be retained in the segregated cluster core. 
Therefore, the models with lower compactness are more representative of the BH population in the local universe, and RI inspirals are more likely to occur once every $\sim$\,300--500 binary-binary encounters.

\subsection{Eccentricity and Prospects of Measurability}

BBH inspirals assembled through dynamical encounters imprint unique features in their eccentricity distributions, which may be a key driver in disentangling the relative rates of various proposed BBH formation scenarios. 
Population modeling predicts highly overlapping distributions of masses that may prove very difficult to leverage in attempts to disentangle formation channels \citep[e.g.,][]{Zevin2017b}, and if natal spins of heavy-stellar-mass BHs are naturally low, as current BBH detections may indicate, the majority of GW spin measurements may also prove uninformative \citep{Farr2017,Farr2017a}. 

In the context of current ground-based GW detectors, GW captures in GC cores are a promising scenario for detectable eccentricity, as the Advanced LIGO/Virgo network will be able to distinguish an eccentric from a circular binary in systems similar to GW150914 if the eccentricity is $\gtrsim$\,0.05 \citep{Lower2018}. 
As seen in Figure \ref{fig:eccentricity}, the eccentricity distribution of GW captures peaks at approximately 0.05 at a GW frequency of 10 Hz, indicating a substantial fraction of these systems will have discernible eccentricity if they are detected. 
Furthermore, we see a spike in the eccentricity distribution of GW captures near $e \approx 1$ from systems that become bound BBHs \textit{inside} the LIGO/Virgo band. 
However, the detectability and selection biases inherent to such highly eccentric sources are difficult to ascertain, as substantial eccentricity will also limit the effectiveness of current matched-filtering searches, which search the data using quasi-circular, aligned-spin templates. 

Ejected and in-cluster binary inspirals will have eccentricities too low to differentiate between circular signals at a GW frequency of 10 Hz. 
However, space-based GW detectors, such as LISA, will be sensitive to orbital frequencies ranging from $10^{-4}$--$10^{-1}$ and orbital eccentricities at 10$^{-2}$ and possibly as low as 10$^{-3}$ \citep{Breivik2016a,Nishizawa2016}. 
We show the eccentricity distribution of these two populations at 10$^{-2}$ Hz with dashed lines in Figure \ref{fig:eccentricity}. 
Similar to the binary-single interactions studied in \cite{Samsing2018} and \cite{DOrazio2018}, we find these populations of BBHs formed through binary-binary encounters to have eccentricities measurable by LISA. 
Therefore, the combination of ground-based and space-based detectors may be useful in disentangling these three dynamically induced inspiral scenarios. 

We also show inspiral times of the three populations in the right panel of Figure \ref{fig:eccentricity}. 
At first glance, the extremely short inspiral timescales of highly eccentric binaries seems promising; the probability of a detection generally scales inversely with the delay time \textit{if} the rate of such interactions is constant throughout the age of the universe. 
On the other hand, these rapid inspiral timescales may cause the majority of such systems to merge early on in the history of the cluster. 
If this is the case, BBH mergers from RIs would occur at redshifts of $z\approx$1--2, above the horizon of current GW detectors. 
However, \cite{Rodriguez2018b} shows that, though ejected BBHs typically merge later due to the large inspiral times between ejection and merger, many BH systems are still retained in GC cores at the present day, and inspirals from GW captures still constitute $\sim$\,10\% of the BH mergers from GCs in the local universe. 

For the most part, the eccentricity distributions of our GC inspirals are consistent with previous work \citep[cf.][]{Samsing2017c, Rodriguez2018b}. 
We note that \cite{Rodriguez2018b} reported a slightly different distribution for the in-cluster binary inspiral population; in Figure \ref{fig:eccentricity} we do not resolve a peak at an eccentricity of 10$^{-3}$. 
This discrepancy is due to a misclassification of triple systems in \cite{Rodriguez2018b} and is resolved in \cite{Rodriguez2018c}. 
We also find that GW captures constitute roughly 17\% of all GC inspirals that originate from binary-binary BH interactions. 
However, this number should taken with caution, as we do not weight our cluster models by the cluster mass function of the local universe, and a proper local rate estimate will need to convolve the formation time of the different GC inspiral populations with their respective inspiral time distribution. 

Because the scattering experiments performed in this study only record the orbital properties of inspiraling binaries once the simulations terminate, the inspiral times of GW captures (i.e., from the IMS binary formation to merger) are approximated by assuming all systems are formed at a reference eccentricity of $(1-e_{\rm ref})=10^{-3}$ (see Section \ref{subsec:inspiral_populations}). 
In the high-eccentricity limit, the inspiral time goes as 
\begin{equation}
\tau_{\rm insp}(a_0,e_0) \propto a_0^4 (1-e_0)^{7/2} \simeq \frac{1}{\sqrt{1-e_0^2}}.\ \ \ \ (1-e_0^2 \ll 1)
\end{equation}
This approximation is supported by the continuous distribution found between the GW capture and in-cluster binary inspiral populations in the right panel of Figure \ref{fig:eccentricity}, where the orbital properties for the latter population are recorded post-interaction and inspiral times calculated in the typical way \citep{Peters1964}. 
Though inspirals near the LIGO/Virgo band may have formed at eccentricities higher than our reference eccentricity, this approximation only constitutes an estimate of the inspiral time at binary formation and does not affect any of the main results in this study. 

Finally, we find that the eccentricity distributions of BBH mergers from binary-binary and binary-single encounters are virtually identical, indicating that the eccentricities themselves will not help decipher which type of resonant dynamical encounter led to the merger. 
Furthermore, the mass distributions of BHs involved in GW captures, in-cluster binary inspirals, and ejected inspirals are indistinguishable. 
However, our scattering experiments weight all interactions over the cluster lifetime equally; properly accounting for cluster evolution and convolving inspiral times with binary formation times may find distinguishing characteristics in the mass distributions for cluster mergers in the local universe \citep[see][]{Rodriguez2018c}. 
Accurate measurements of the rate of highly eccentric BBH mergers from an accumulation of GW detections over next few years will help to establish a rate of these exotic signals and provide further insight into the relative contribution of binary-binary interactions in facilitating GW captures during RIs within GCs. 
Furthermore, if the rate of GW captures compared to in-cluster binary inspirals proves to be constant across clusters, we could leverage the detection of highly eccentric signals to gain insight into the total rate of BBH mergers from GCs.

\subsection{Post-Newtonian Equations of Motion}
We find the inclusion of pN terms in $N$-body scattering experiments to have a negligible effect on the Newtonian endstates of binary-binary encounters. 
In the context of GW captures, previous studies have already discovered that the energy-dissipative 2.5pN term plays an important role in facilitating in-cluster mergers in hardened, eccentric binaries between encounters and in the chaotic RIs themselves. 
This study was the first to examine the effect of lower-order, non-dissipative 1pN and 2pN terms in the equations of motion used in $N$-body scattering experiments. 
Though the 1pN and 2pN terms do not dissipate orbital energy, these lower-order terms are the primary driver of certain aspects of general relativistic orbital evolution, such as periapse precession, and play an important role in the stability of secularly evolving systems, such as hierarchical triples \citep{Blaes2002}. 

Particularly in the case of four-body encounters, the inclusion of these terms may have proven important in accurately capturing the probability of GW captures, especially if many inspirals were the result of short-lived hierarchical triples. 
However, we find that these terms have no noticeable effect on the probability of GW captures during resonating encounters, implying that using only the 2.5pN term suffices to accurately capture the probability of inspirals during strong BH encounters in GCs. 

Nonetheless, the strong encounters examined in this study are in a highly relativistic regime; as BBHs approach merger, their velocities reach appreciable fractions of the speed of light. 
Though higher-order pN terms will not affect our probabilistic measurements, truncating the pN expansion may lead to inaccuracies in the integration of the system as it approaches merger --- and thereby lead to errors in the measurements of orbital quantities, such as the SMA and eccentricity at a particular GW frequency. 
Therefore, we stop our integration at a particular threshold value of the SMA, namely when the binary SMA is less than ten times the sum of the two BH Schwarzschild radii (equivalent to 40$M$ in geometrized units, assuming equal masses). 
Two 20 $M_\odot$ BHs on a circular orbit at this SMA would be moving at $\sim$\,0.2$c$, meaning the contribution from the 3pN term is $\sim$\,0.2\% that of the lowest order pN term. 
Furthermore, such a system would merge in just a few seconds, making the possibility of perturbations from other components in the encounter negligible. 
However, terminating the integration at larger SMA values should still suffice; even at an SMA of 100$M$, two 20 $M_\odot$ BHs on a circular orbit would merge in less than 100\,s. 
In the future, when performing pN $N$-body scattering experiments, it may therefore be more accurate to halt the $N$-body integration at larger orbital separations and evolve the orbital properties of the system forward numerically to the GW frequencies of interest. 
This can either be accomplished by terminating the simulations once an assigned tidal threshold is surpassed, as in \cite{Samsing2014}, or by choosing a fixed orbital separation at which to terminate --- one large enough that the Newtonian orbital parameters are still accurate, yet small enough that perturbations from other particles in the interaction are negligible for the remainder of the inspiral; this methodology is explored further in \cite{Rodriguez2018c}.

\section{Conclusions}\label{sec:conclusions}

In this study, we have systematically investigated the contribution of strong binary-binary encounters to the population of eccentric BH inspirals in GCs, derived scaling relations for GW captures due to binary-binary interactions, quantified the importance of lower-order (non-dissipative) pN terms in facilitating eccentric BBH mergers, and gauged the efficiency and properties of eccentric GW captures from realistic cluster models. 
Our key findings are: 

\begin{enumerate}
\item
Though less common than binary-single BH interactions in GCs, binary-binary BH interactions are $\sim$\,5 times more likely to induce a GW capture during a RI, and therefore contribute to $\sim$\,25--45\% of the total number of highly eccentric inspirals originating from strong encounters in GCs, where the remaining are mostly from binary-single interactions.

\item
The GW capture probability from binary-binary encounters follows the same SMA and mass scaling relation as binary-single encounters: ${\rm \mathcal{P}_{\rm cap}} \propto (M/a_0)^{5/7}$.

\item
GW capture probabilities for both binary-binary and binary-single interactions monotonically increase as a function of the compactness of the cluster environment, scaling linearly with compactness for moderately and highly compact GCs. 

\item
The relative contribution of GW captures induced from binary-binary interactions to the total number of GW captures is mildly sensitive to the cluster compactness, with the contribution of GW captures from binary-binary encounters being slightly less prominent in the most compact clusters. 

\item
Non-dissipative pN terms play a negligible effect in inducing inspirals during chaotic BH encounters in GCs; the GW capture cross section can be accurately captured by only including the Newtonian and 2.5pN terms in the $N$-body equations of motion. 

\item
GW inspirals produced from binary-binary encounters in GCs are similar to those produced from binary-single encounters and lead to three distinct populations of BBH mergers: ejected inspirals, in-cluster binary inspirals, and GW captures. 
The BBH mergers from each population have a distinct eccentricity distribution. 
GW captures generally have eccentricities measurable by LIGO/Virgo, whereas in-cluster binary mergers and ejected mergers have eccentricities measurable by LISA. 

\item
Eccentric BBH inspirals formed in the cores of GCs occur at rates accessible to the Advanced LIGO/Virgo network at design sensitivity.
A single observation of such a signal will highly constrain its formation scenario, and a population of such detections could lead to the most stringent constraints on the relative rates of BBH formation channels. 

\end{enumerate}

\acknowledgments
M.Z. would like to express thanks to Kyle Kremer, Chris Pankow, and Fred Rasio for stimulating discussions and valuable insight into this study, and Christopher Berry for useful comments on this manuscript. This work was initiated and supported by the 2017 Kavli Summer Program in Astrophysics at the Niels Bohr Institute in Copenhagen, and the authors would like to thank DARK at the University of Copenhagen for incredible hospitality. The 2017 Kavli Summer Program program was supported by the Kavli Foundation, Danish National Research Foundation (DNRF), the Niels Bohr International  Academy and  DARK.   M.Z. greatly appreciates support from the NSF GK-12 graduate student fellowship, under grant No. DGE-1007911.  C.R. acknowledges support from the Pappalardo Fellowship in Physics at MIT. E.R. thanks the DNRF for support as a Niels Bohr Professor.


\bibliographystyle{yahapj}
\bibliography{library}


\end{document}